# Title: Topology and Dynamics of Active Nematic Vesicles

**Authors:** Felix C. Keber[1,3,†], Etienne Loiseau[1,†], Tim Sanchez[2,†], Stephen J. DeCamp[2], Luca Giomi[4], Mark J. Bowick[5], M. Cristina Marchetti[5], Zvonimir Dogic[2,3] and Andreas R. Bausch[1]

**Affiliations:**

[1] Department of Physics, Technische Universität München, 85748 Garching, Germany.

[2] Department of Physics, Brandeis University, Waltham, MA 02474, USA.

[3] Institute for Advanced Study, Technische Universität München, 85748, Garching, Germany.

[4] SISSA International School for Advanced Studies, Via Bonomea 265, 34136 Trieste, Italy.

[5] Physics Department and Syracuse Biomaterials Institute, Syracuse University, Syracuse, NY 13244, USA.

[†] These authors contributed equally to this work.

**Abstract:** Engineering synthetic materials that mimic the remarkable complexity of living organisms is a fundamental challenge in science and technology. We study the spatiotemporal patterns that emerge when an active nematic film of microtubules and molecular motors is encapsulated within a shape-changing lipid vesicle. Unlike in equilibrium systems, where defects are largely static structures, in active nematics defects move spontaneously and can be described as self-propelled particles. The combination of activity, topological constraints and vesicle deformability produces a myriad of dynamical states. We highlight two dynamical modes: a tunable periodic state that oscillates between two defect configurations, and shape-changing vesicles with streaming filopodia-like protrusions. These results demonstrate how biomimetic materials can be obtained when topological constraints are used to control the non-equilibrium dynamics of active matter.

**One sentence summary**: Covering a vesicle with microtubule based active liquid crystals powered by molecular motors produces tunable clocks and shape-changing life-like materials.

**Main text:** Fundamental topological laws prove that it is not possible to wrap a curved surface with lines without encountering at least one singular point where the line is ill-defined. This mathematical result is familiar from everyday experience. Common examples are decorating the earth's surface with lines of longitude or latitude or covering a human hand with parallel papillary ridges (fingerprints). Both require the formation of singular points known as topological defects (*1*). The same mathematical considerations apply when assembling materials on microscopic length scales. Nematic liquid crystals are materials whose constituent rod-like molecules align spontaneously along a preferred orientation that is locally described by the director (line) field. Covering a sphere with a nematic leads to the formation of topological defects called disclinations. Mathematics dictates that the net topological charge of all defects on a spherical nematic has to add up to +2, where a charge of *s* denotes a defect that rotates the director field by $2\pi s$. The basic nematic defects have charge +1/2 or -1/2, corresponding to a $\pi$ rotation of the director field in the same, or the opposite sense, respectively, as that of any closed path encircling the defect (*2-6*). Defects on topologically constrained geometries can acquire highly complex and tunable spatial arrangements, which can drive assembly of intriguing higher order hierarchical materials (*7, 8*). For this reason combining conventional soft materials with topological constraints has emerged as a promising platform for organizing matter on micron length scales. So far the majority of studies in this area has focused on investigating equilibrium materials confined on rigid non-deformable surfaces of varying topology (*9-17*). Equilibrium statistical mechanics imposes, however, tight constraints on properties of such topological soft materials, which can acquire remarkable spatial complexity but cannot sustain any persistent temporal dynamics.

Recent advances have enabled assembly of active nematic liquids crystals in which the constituent rod-like building blocks continuously convert chemical energy into mechanical motion (*18-20*). Such materials are no longer constrained by laws of equilibrium statistical mechanics. Consequently, unconfined active nematics exhibit highly sought-after properties such as spontaneous chaotic flows that are tightly coupled to continuously generating and annihilating motile defects (*21-23*). We have merged active nematics with soft topological constraints to create topological active matter. Microtubule based active nematics were confined onto a deformable spherical surface provided by a lipid vesicle. Similar to well-studied equilibrium nematics confined on a sphere, topology requires formation of defects with net topological charge of +2. In stark contrast to the equilibrium case, however, activity generated by energy-consuming kinesin motors endows the active nematic defects with motility. As a result, the complex spatial defect structure become dynamic and the active nematic vesicle turned into robust colloidal clocks with tunable frequency. By controlling the vesicle tension we coupled the oscillatory dynamics of the active nematic cortex to vesicle deformations, creating biomimetic shape-changing materials. Our experiments suggest a route for designing soft materials with controlled oscillatory dynamics. It further raises intriguing questions about how the dynamics of topological active matter can be enriched by increasing the complexity of the constraining surface, by controlling the nature of the synchronization transitions that occur in arrays of colloidal oscillators, and by utilizing other far-from-equilibrium materials such as active crystals (*24*).

**Assembly of active nematic vesicles:** We build on recent work by encapsulating active nematics into deformable vesicles. Active nematic vesicles are produced by encapsulating microtubules, kinesin motors clusters and non-adsorbing polymer polyethylene glycol (PEG) within a lipid vesicle using emulsion transfer technique (*25*). PEG induces adsorption of microtubule filaments onto the inner leaflet of the vesicle by the depletion mechanism *(26)*. At high microtubule concentration the entire vesicle surface is coated with a dense microtubule monolayer of extensile bundles, effectively creating a two-dimensional (2D) nematic cortex. Individual kinesin motors, fueled by energy from adenosine triphosphate (ATP) hydrolysis, processively move along a microtubule backbone at velocities up to 0.8 μm/s (*27*). When bound into multi-motor clusters through a biotin-streptavidin linkage, kinesin clusters crosslink adjacent microtubules inducing their relative sliding and generating active extensile stresses (*19, 28, 29*). We image active nematic vesicles by confocal microscopy (*30*).

In equilibrium there are multiple defect arrangements that minimize the free-energy of a 2D spherical nematic, with the exact configuration depending on the strength of the elastic constants. In the one elastic constant approximation, the free energy is minimized when four +½ defects are located at the corners of a tetrahedron inscribed within the sphere (*2, 3, 5*). This configuration is favored because defects of the same charge repel each other. Placing them at the corners of a tetrahedron maximizes their separation, thus minimizing liquid crystal distortions. The 3D reconstruction of the surface bound active nematic reveals the presence of four +½ disclination defects (Fig. 1), in agreement with theoretical predictions for equilibrium systems and previous experimental observations of spherical nematic shells of finite thickness (*4*).

**Oscillating defect dynamics in active nematic vesicles:** At finite ATP concentrations, active energy input provided by kinesin clusters drives the microtubule nematic far-from-equilibrium, yielding surprising dynamics. In planar nematics active stresses destabilize the homogeneous state (*31-33*) and generate self-sustained streaming flows, with the continuous creation and annihilation of motile defects that interact through elastic and hydrodynamic forces (*19, 21-23*). When the nematic film is confined to the surface of a sphere, active unbalanced stresses drive motility of the four +½ surface bound disclinations, which leads to streaming flows of the entire vesicle (Fig. 1, Movie S1). The dynamics of spherical active nematics is simpler than that of their planar counterpart, as the four +½ defects on spherical nematics never disappear and their lifetime is limited only by the vesicle stability. Furthermore, the spatial confinement used in our experiments suppresses the production of additional defect pairs, thus providing a unique opportunity to study the dynamics of a few isolated interacting defects in a controlled way. We speculate that the defect proliferation seen in planar nematics does not occur in our vesicles because they are too small. In all cases, the vesicles' diameter is well below the length scale, $\ell_a$, at which the homogeneously ordered system is unstable to bend deformations. For MT based planar active nematics $\ell_a$ is estimated to be ~100 μm (*19*).

Similar to the equilibrium case, the repulsive elastic interactions between four +½ disclinations in an active spherical nematic favor a tetrahedral defect configuration, *(2, 3, 5)*. In active systems, however, the asymmetric shape of comet-like +½ disclinations also generates active stresses and associated flows that in turn drive defect motion. For extensile systems defects are propelled at constant speed towards the head of the comet (Fig. 1C) (*21*). It is not possible for the four defects to simultaneously minimize elastic repulsive interactions and move with prescribed velocity determined by ATP concentration while keeping their relative distance constant. As a result, defects move along complex spatiotemporal trajectories. To elucidate this emergent dynamics we imaged the time evolution of active vesicles using confocal microscopy and traced the 3D position of the individual defects (Fig. 1, Movie S1). At any given time the positions of the four defects are described by the variables $\alpha_{ij}$, which denote the angle between radii from the vesicle center to each of the 6 defect pairs, ij (Fig. 2A). For a tetrahedral configuration all six angles are $\alpha_{ij}$=109.5°, while for a planar configuration $\alpha_{12}=\alpha_{23}=\alpha_{34}=\alpha_{41}=90°$ and $\alpha_{13}=\alpha_{24}=180°$ (and permutations), resulting in an average angle of $\langle \alpha \rangle_{planar} = \frac{1}{6}\sum_{i<j=1}^{4} \alpha_{ij} = 120°$ (Fig. 2A). The temporal evolution of all six angles reveals a clear pattern of defect motion (Fig. 2B). For example, at time t=602 seconds two angles assume a large value near 180°, while the other four are approximately 90°, indicating a planar configuration. Forty-three seconds later this configuration switches to a tetrahedral configuration in which all angles are equal (Fig. 2D). Observations on longer timescales demonstrate that the defects repeatedly oscillate between the tetrahedral and planar configurations, with a well-defined characteristic frequency of 12 mHz (Fig. 2C). The frequency is set by the motor speed and the size of the sphere, and can be tuned by the ATP concentration, which determines the kinesin velocity (Fig. S1) *(27)*.

**Particle-based theoretical model describes oscillatory dynamics of active nematic vesicles**: The oscillatory dynamics of spherical nematics can be described by a coarse-grained theoretical model. As shown recently, +½ defect in extensile nematics behave as a self-propelled particles with velocity $v_0$ proportional to activity directed along the axis of symmetry, $\boldsymbol{u}$ (Fig. 3A, Fig. S2) *(10)*. Each defect is then characterized by a position vector on the sphere $\boldsymbol{r}_i = \boldsymbol{r}_i(\theta, \varphi)$, where $\theta$ and $\varphi$ are the spherical coordinates, and a unit vector $\boldsymbol{u}_i = (\cos\psi_i, \sin\psi_i)$ describing the orientation of the comet axis and directed from the tail to the head. In local coordinates $\boldsymbol{u}_i = \cos\psi_i\, \boldsymbol{e}_{\theta i} + \sin\psi_i\, \boldsymbol{e}_{\phi i}$, where $\boldsymbol{e}_{\theta i}$ and $\boldsymbol{e}_{\phi i}$ are unit vectors in the latitudinal and longitudinal directions and $\psi_i$ the local orientation (Fig. S3). Adapting the planar translational dynamics to the curved surface of the sphere, and augmenting it with the dynamics of orientation, the equations of motion of each defect are given by the overdamped Newton-Euler equations for a rigid body

$$\zeta_t \left(\frac{d\boldsymbol{r}_i}{dt} - v_0 \boldsymbol{u}_i\right) = \boldsymbol{f}_i, \qquad \zeta_r \frac{d\psi_i}{dt} = M_i, \tag{1}$$

where $\boldsymbol{f}_i = -dE/d\boldsymbol{r}_i$ and $M_i = -dE/d\psi_i$, with $E$ the elastic energy of the defects, are the force and torque on the *i*-th defect due to the repulsive interactions with the other three, and $\zeta_t$ and $\zeta_r$ are translational and rotational frictions, respectively *(30)*. The dynamical equations are solved numerically on the sphere, with defects initially randomly spaced along the equator and with random orientations. In equilibrium, $v_0 = 0$ and the defects relax toward the minimum of their potential energy, with the four defects sitting at the vertices of a regular tetrahedron (Fig. 3B, Fig. S4).

In the presence of activity, a finite $v_0$ allows the defects to escape from their minimal energy configuration. This leads to an oscillatory dynamics characterized by a periodic motion of the defects between two symmetric tetrahedral configurations (Figs. 3C, E), passing through an intermediate planar one (Fig. 3D). The oscillations arise from the competition between the active force $\zeta_t v_0 \boldsymbol{u}_i$ and the elastic force $\boldsymbol{f}_i$. The period is determined by the sum of the time scales for the defect to move uphill and downhill in the energy landscape (Fig. 3F, Fig. S5, Movie S2). The period of the oscillations is equal to the time required for a defect to perform a full revolution around the sphere axis; thus $T \approx 2\pi R/v_0$ and the frequency increases linearly with activity (Fig. S6). Oscillations occur provided the rotational dynamics is fast relative to the translational dynamics. For the defects to overshoot when they approach the minimum in the energy landscape, they need to quickly reorient before being attracted back into the energy minimum (Fig. S7). The dynamics of the average angle $\langle \alpha \rangle$ (Fig. 3F) reproduces very closely our experimental findings (Fig. 2B).

The perpetual oscillatory dynamics of active nematic vesicles results thus from the interplay of topology, liquid crystalline-order and activity. The spherical topology of the vesicles imposes the existence of exactly four +½ defects with particle-like features. The entropic elasticity originating from the local nematic order shapes the energy landscape through which the defects move. Finally, activity powered by ATP hydrolysis fuels the defect motion. The outlined mechanism leads to robust oscillatory dynamics that survives the presence of noise that is inevitable in the experimental system.

**Active nematic cortex drives large-scale vesicle shape changes:** The flexibility of the encapsulating vesicle allows us to study the uncharted regime in which the dynamics of an active nematic cortex couples to a deformable constraining surface (Movie S3). To explore the dynamics of this entire regime we apply a hypertonic stress inducing a water efflux that deflates the vesicle (Fig. 4, Movie S4). The shape of slightly deflated vesicles continuously fluctuates around a mean spherical shape and is characterized by the continuous growth and shrinkage of the major and minor axis of an ellipse, with a periodicity set by the defect transport speed. In addition, these vesicles exhibit four motile protrusions that are tightly coupled to the dynamics of the underlying disclination defects. Deflating the vesicles further causes a dramatic shape change: the overall vesicle becomes anisotropic and motile, with filopodia-like protrusions growing in size and reaching lengths of tens of microns. The 3D reconstruction of shape-changing flaccid vesicle demonstrates the existence of only four protrusions (Fig. 4B), this number being determined by the fundamental topological constraints. Re-swelling the vesicle causes the amplitude of the shape deformations to continuously decrease, the protrusions to vanish and the vesicle to recover its initial spherical shape.

The other parameter that critically affects the emergent behavior of spherical nematics is the vesicle diameter *(34, 35)* (Fig. 5). Decreasing the diameter increases the curvature of surface bound microtubules and the energetic cost of confining active nematics to a spherical surface. For vesicles with radii larger than 18 μm we only observe oscillations of four motile +½ defects (Fig. 5H). For smaller radii, the population of vesicles and the emergent dynamics becomes heterogeneous (Fig. 5I). For example, we observe a dynamical mode in which microtubules form a rotating ring around the equator of a spherical vesicle (Fig. 5A). The diameter of the microtubule ring increases due to extensile forces driven by molecular motors. Eventually, to satisfy the spherical confinement, the ring buckles out of the equatorial plane forming a saddle shape configuration which initiates the formation of the four +½ defects. The defects collide pairwise and fuse such that a new ring is formed (Fig. 5C and D, Movie S5, S6). By this mechanism the vesicle switches between the ring structure and the structure characterized by four +½ defects (Fig. 5B). For the same parameters, we also find that stiff microtubules can deform the vesicle, forming a spindle like structure with two +1 defects at the spindle poles. Structurally, spindle-like vesicles resemble the isotropic-nematic tactoids found in equilibrium liquid crystals (Fig. 5E) *(36)*. Active spindle-like vesicles also exhibit interesting dynamics. Taking advantage of the excess membrane area stored in the membrane tension the constituent extensile microtubule bundles push outwardly to extend the spindle poles. At a critical bundle length, the connectivity of the constituent microtubule bundles in the central spindle region decreases and the spindle becomes unstable, buckles and folds into a more compact shape. At this point the microtubules start extending again and the cycle of vesicle extension, buckling, folding and re-extension repeats multiple times (Fig. 5F, G, Movie S7). It is also possible to observe vesicles that transition between spindle and ring structures (Fig. S8, Movie S8).

In conclusion, ours and other recent advances demonstrate how combining active matter with topological constraints yields structures and dynamics that cannot be observed in conventional equilibrium systems (*37-40*). Specifically, we have shown that confining active nematic films onto the surface of a sphere suppresses defect generation and yields a robust and tunable oscillatory dynamics that is well described by a coarse-grained theoretical model. Furthermore, confinement and the constraints imposed by topology significantly simplify the complex dynamics of planar active nematics that exhibit complex chaotic spatiotemporal behavior.

**Acknowledgments:** Research was supported by ERC-SelfOrg (FCK, EL, ARB), partly by the SFB863 and the Nanosystems Initiative Munich (FCK, EL, ARB), the Institute of Advanced Study-Technical University Munich (ZD, FCK), the NSF under awards DMR-1305184 and DGE-1068780 (MCM), and the Soft Matter Program at Syracuse University (MJB, MCM). The primary support for work at Brandeis was provided by the W. M. Keck Foundation and NSF-MRSEC-0820492 (TS, ZD). Acquisition of ATP dependence data was supported by DOE-BES under Award DE-SC0010432TDD (SJD, ZD). We also acknowledge use of Brandeis MRSEC optical microscopy facility.

# Figures

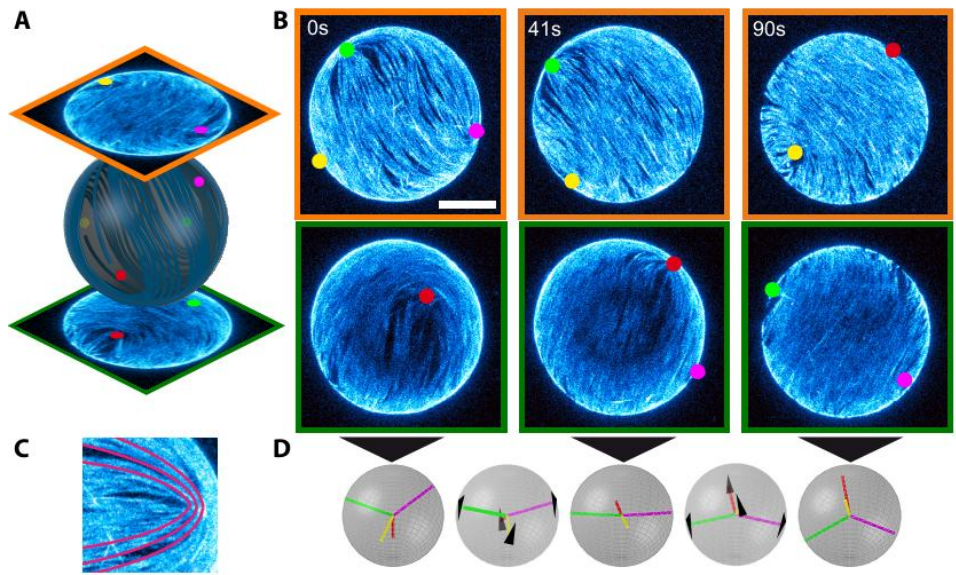

**Figure 1: Defects dynamics of an active nematic film on the surface of a spherical vesicle: (A)** Hemisphere projection of a 3D confocal stack of a nematic vesicle identifying the position of four +½ disclination defects. **(B)** Time series of hemisphere projections over a single period of oscillation in which the four defects switch from tetrahedral (t=0s) through planar (t=41s) and back to tetrahedral (t=90s) configurations. Bar is 20μm. **(C)** Comet-like +½ disclination defect with schematic of the orientation of the nematic director (red lines). **(D)** Schematic of the defect configurations at the timepoints of (B) and intermediate times (t=24s, t=65s). The black arrows indicate the direction of defect motion.

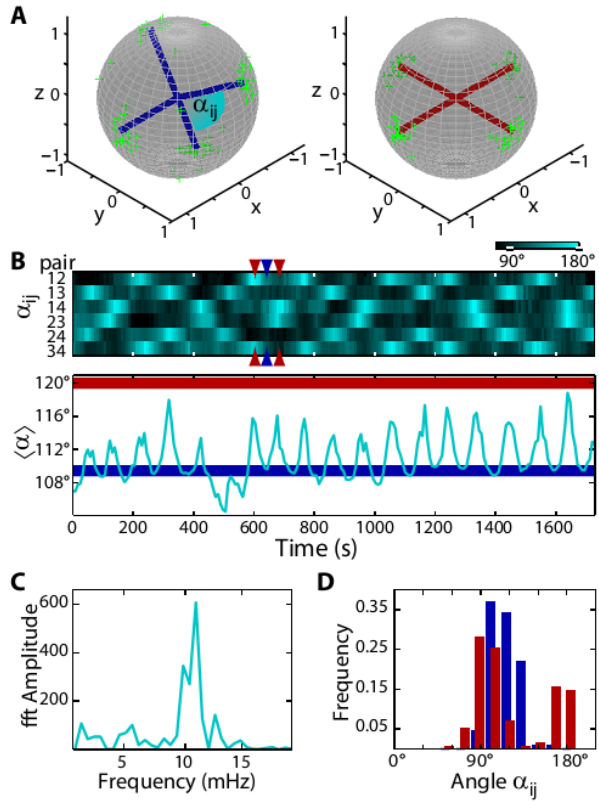

**Figure 2: Oscillatory dynamics of topological defects.** (**A**) Tetrahedral (blue) and planar (red) defect configurations. The green scatter plots show the measured positions of the defects on the unit sphere at the extremal configurations. (**B**) Top: Kymograph showing the time evolution of angular distances α$_{ij}$ of all six defect pairs (as indicated in A). Bottom: The average angle oscillates between the tetrahedral configuration ($\langle \alpha \rangle$ = 109.5°, blue line) and the planar configuration ($\langle \alpha \rangle$ = 120°, red line). An exemplary transition between the two configurations is indicated by the colored arrows (t= 602s, t= 643s, t= 684s). (**C**) Power spectrum of $\langle \alpha \rangle$. The peak at 12 mHz is associated with tetrahedral-planar oscillations. (**D**) Distributions of angles α$_{ij}$. Gaussian fits return angles of 109° ± 13° for the tetrahedral configuration (blue) and for the planar configuration (red) 90° ± 12° and 163° ± 9°.

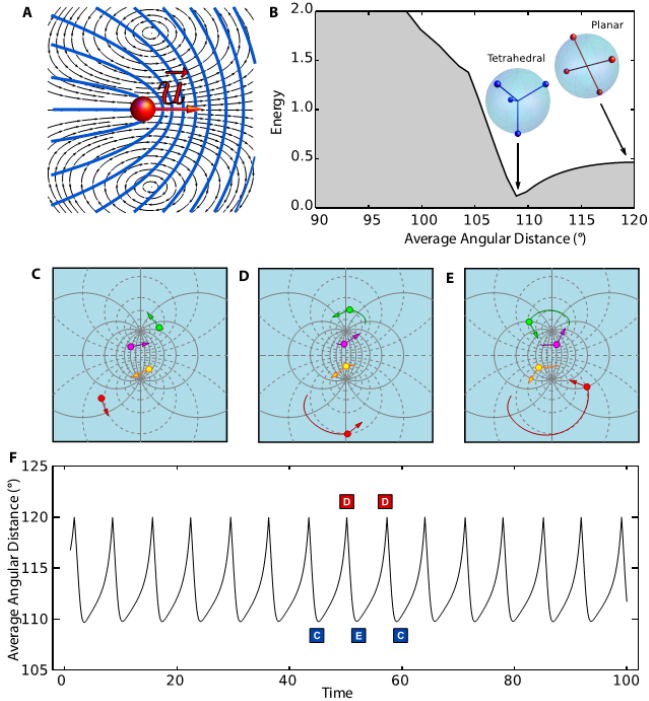

**Figure 3**: **A theoretical model that describes nematic defects as self-propelled particles predicts oscillatory dynamics.** **(A)** Active +½ disclinations with an axis of symmetry ***u*** behave as self-propelled particles by generating a local flow (black lines) that convects the defect core toward the head of the comet-like structure described by the director field (blue lines). **(B)** In the absence of activity four +½ sphere-bound disclinations relax toward the minimum of their potential energy, with the four defects sitting at the vertices of a regular tetrahedron. For active nematics the defects undergo a self-organized periodic motion: starting from a passive equilibrium tetrahedral configuration **(C)** they pass through a planar configuration **(D)** on the way to another tetrahedral configuration **(E)** and then back again periodically. **(F)** Images C-E show the stereographic projection of the defect configurations in the plane. The average angular distance $\langle \alpha \rangle$ as a function of time with asymmetric oscillations between a tetrahedral state (i.e. $\langle \alpha \rangle = 109.5°$) and a planar state (i.e. $\langle \alpha \rangle = 120°$). The energy landscape reveals a minimum for the tetrahedral configuration and a maximum for the planar one. The average angle $\langle \alpha \rangle$ does not distinguish between the two equivalent tetrahedra C and E. When described in terms of this coordinate, the dynamics oscillates from the minimum of the plot in frame B to the maximum, then back to the same minimum.

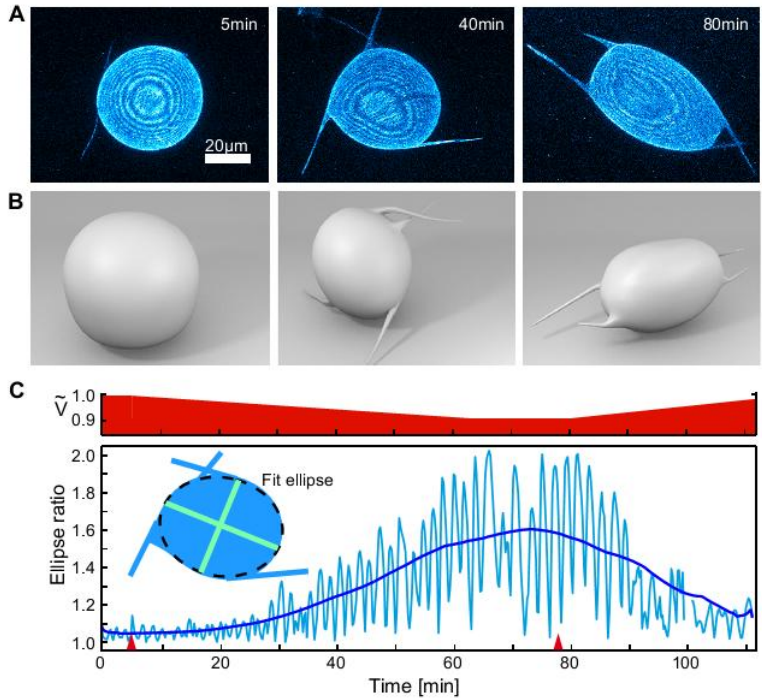

**Figure 4: Vesicle shape changes driven by defect dynamics.** A 10% hypertonic stress is applied at t=5.2 min to provide an excess membrane area. (**A**) Confocal images showing the z-projection of the vesicle shape, with corresponding 3D schematics shown in (**B**). Starting from a spherical nematic vesicle with four +½ defects (t=5min), four dynamic protrusions grow from the defect sites while the vesicle slowly deswells. At t=78 min the vesicles reswells and the protrusions decrease in size and eventually disappear. (**C**) Bottom: The amplitude of shape deformations increases over time, as shown by the plot of the vesicle aspect ratio (major to minor ellipse axis), reaching a maximum value of 1.6 at t=75 min. Restoring $\tilde{V}$ by applying a hypotonic stress at t=78 min inverts the effect, suppressing shape fluctuations. Red arrows denote when osmotic stresses are applied. Top: Estimate of the time evolution of the vesicle reduced vesicle volume $\tilde{V}$.

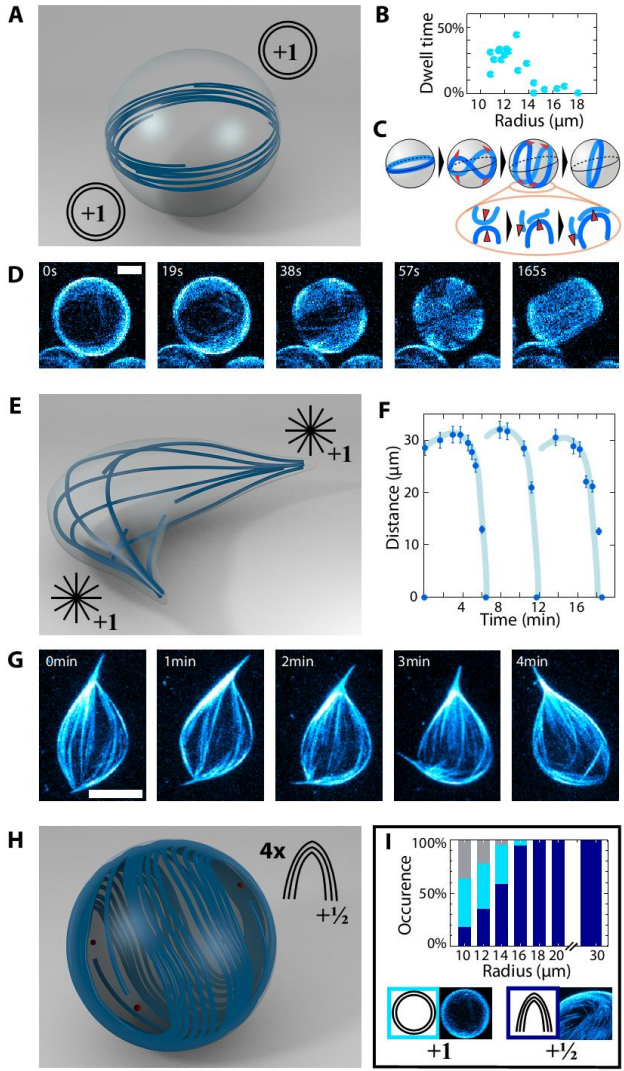

**Figure 5: Size dependent morphology and dynamics of flexible nematic vesicles.** (**A-D**) Ring-mode vesicle. (**A**) Microtubule bundles arrange in a ring around the equator. (**B**) Fraction of time spent by the vesicles in the ring configuration. (**C**) Schematic illustrating the transition between the ring and the four +½ defect configuration. The corresponding z-projected confocal images to this sequence are shown on (**D**). (**E-G**) Spindle-like vesicle. (**E**) Each pole contains a +1 aster-like defect. (**F**) Temporal evolution of the distance between the two +1 aster-like defects. While extending, the microtubule bundles buckle and the two +1 protrusions fold on each other. This results in cycles of microtubule extension, buckling and folding (light blue lines as guide to the eyes). A sequence of confocal images illustrating this dynamics is shown on (**G**). (**H**) Schematic of a large spherical vesicle exhibiting four +½ defects. (**I**) Types of dynamics: the histogram shows the percentage of vesicles displaying a defect configuration in dependence of the radius (total count =168). For radii above 18 μm, all the vesicles exhibit the four +½ defect configuration shown in **H**. For radii in the range 10-18 μm vesicles which undergo continuous transitions between four +½ defect (blue) and ring (cyan) configurations (shown in A) are found. The colors code for the defect topology as shown in the pictograms, some cases remain uncharacterized due to resolution limitations. Bars are 8μm.